# E-beam-enhanced solid-state mechanical amorphization of α-quartz: Reducing deformation barrier via localized excess electrons as mobile anions

Sung-Gyu Kang[1,+], Wonseok Jeong[2,+], Hwangsun Kim[1], Jeongin Paeng[1], Seungwu Han[1], and Heung Nam Han[1,*], In-Suk Choi[1,*]

[1] Department of Materials Science and Engineering & Research Institute of Advanced Materials, Seoul National University, Seoul, Republic of Korea

[2] Lawrence Livermore National Laboratory, Livermore, USA

* Corresponding authors: Prof. Heung Nam Han and Prof. In-Suk Choi

E-mail: hnhan@snu.ac.kr and insukchoi@snu.ac.kr

[+] These authors equally contributed to this work.

## Abstract

Under hydrostatic pressure, α-quartz (α-SiO$_2$) undergoes solid-state mechanical amorphization wherein the interpenetration of [SiO$_4$]$^{2-}$ tetrahedra occurs and the material loses crystallinity. This phase transformation requires a high hydrostatic pressure of 14 GPa because the repulsive forces resulting from the ionic nature of the Si–O bonds prevent the severe distortion of the atomic configuration. Herein, we experimentally and computationally demonstrate that e-beam irradiation changes the nature of the interatomic bonds in α-quartz and enhances the solid-state mechanical amorphization at nanoscale. Specifically, during *in situ* uniaxial compression, a larger permanent deformation occurs in α-quartz micropillars compressed during e-beam irradiation than in those without e-beam irradiation. Microstructural analysis reveals that the large permanent deformation under e-beam irradiation originates from the enhanced mechanical amorphization of α-quartz and the subsequent viscoplastic deformation of the amorphized region. Further, atomic-scale simulations suggest that the delocalized excess electrons introduced by e-beam irradiation move to highly distorted atomic configurations and alleviate the repulsive force, thus reducing the barrier to the solid-state mechanical amorphization. These findings deepen our understanding of electron–matter interactions and can be extended to new glass forming and processing technologies at nano- and microscale.

# 1. Introduction

Can we alter the mechanical properties of nanomaterials athermally? Recently, the irradiation of nanomaterials with photons or electrons has been reported to result in immediate changes in their intrinsic mechanical properties during real-time mechanical testing. For example, in the absence of light irradiation, brittle zinc sulfide (ZnS) crystals show plasticity, [1] whereas light irradiation leads to their brittle fracture. This can be explained by the excess charges in the ZnS crystal introduced by photoexcitation that limit dislocation mobility. Under ultraviolet irradiation, the ZnS crystal shows an increase in the elastic modulus and hardness, which can be attributed to the increased electronic strain and Peierls barrier [2]. Under electron beam (e-beam) irradiation, brittle ceramic nanostructures, such as amorphous silica [3–8], amorphous alumina [9], single-crystal silicon [10,11], and single-crystal zinc oxide (ZnO), [12] show beam-dependent plastic deformation. Among these materials, amorphous silica is the most intensively studied, and the viscoplastic deformation of brittle amorphous silica under e-beam irradiation has been investigated under various e-beam conditions. These studies have revealed that, depending on the acceleration voltage, super-plastic deformation is induced via knock-on damage (>200 keV) [4–8] or radiolysis (≤30 keV) [3,13]. Notably, in a previous study, we experimentally demonstrated that the mechanical properties of a selected area of nanosilica could be engineered by modulating the interaction volume of e-beam irradiation. Thereafter, we computationally demonstrated that the excess charges introduced by the e-beam irradiation change the nature of the interatomic bonds, and the volume where the electron–matter interaction occurs undergoes viscoplastic deformation [13]. Similarly, Hirakata *et al.* reported a decrease in the pop-in load during the nanoindentation of single-crystal ZnO under e-beam irradiation [12]. The reduction in the critical resolved shear strength was attributed to the excess electrons and holes, which change the intrinsic bond strength of the material, and this behavior was modeled through atomic-scale simulations. However, the mechanism and role of the excess charges during deformation have not yet been elucidated.

Herein, we first report that e-beam irradiation enhances the mechanical amorphization of a single-crystalline α-quartz (α-SiO$_2$) micropillar at room temperature during *in situ* compression tests. Owing to its sparse atomic structure, α-quartz, a stable crystalline silica polymorph at room temperature, is amorphized under hydrostatic pressure; that is, solid-state amorphization occurs [14–16]. However, this room-temperature amorphization requires an extremely high hydrostatic pressure, approximately 14 GPa, because it results in the severe distortion of the Si–O–Si bond angle, which causes a large repulsive coulombic force between the Si atoms in the [SiO$_4$]$^{2-}$ tetrahedra. Therefore, we hypothesized that the deformation behavior and amorphization of single-crystalline α-quartz will be affected by e-beam irradiation, similar to the occurrence of e-beam-induced viscoplasticity in amorphous silica, because the basic building blocks and type of bonds in crystalline α-quartz are the same as those of amorphous silica, which also comprises a collective network of [SiO$_4$]$^{2-}$ tetrahedra. Hence, we systematically performed *in situ* scanning electron microscopy (SEM) compression tests on α-quartz micropillars and confirmed that the e-beam facilitates the solid-state amorphization of α-quartz. Analysis of the deformation behavior and microstructure revealed that the critical hydrostatic pressure for the solid-state amorphization drastically decreased under e-beam irradiation. Subsequently, the changes in the nature of the atomic bonds and the effects of the excess charges were computationally investigated, revealing that the excess charges enable the close Si–Si and O–O configurations required for the solid-state amorphization and reduce the critical hydrostatic pressure. The computational results suggest that, during deformation, the uniformly distributed excess electrons introduced by the e-beam irradiation move to highly distorted atomic configurations; further, they are located like anions between Si atoms, thus reducing the repulsive forces, acting as network modifiers, and enabling the mechanical solid-state amorphization. Our findings are expected to contribute to the development of new methods for controlling glass formation at small scales without changing the temperature or composition.

## 2. Materials and Methods

*2.1. Sample preparation and in situ SEM mechanical tests*

The α-quartz micropillars were fabricated on a commercial-grade (0001) α-quartz single-crystal substrate using focused-ion-beam milling (FEI Nova 600 NanoLab™). An ion beam with an acceleration voltage of 30 kV was used for both coarse (3 nA) and fine (10 pA) substrate milling. The diameter, height, and taper angle of the pillars were 290 nm, 970 nm, and 2.4°, respectively. Subsequently, we performed displacement-controlled compression tests on the pillars at a constant strain rate of 0.0006 s$^{-1}$ and a maximum strain of 0.15 with a 500-nm-radius flat punch indenter using an *in situ* mechanical test system (Hysitron PI-85 SEM Picoindenter®) installed in a scanning electron microscope (FEI Nova 600 NanoLab™).

*2.2. Microstructural characterization*

Cross sections parallel to the direction of the incident electron beam of the α-quartz pillars before and after compression tests were sampled using focused-ion-beam milling (FEI Nova 600 NanoLab™). Diffraction patterns and ($11\bar{2}0$) dark-field transmission electron microscopy (TEM) images were acquired using a JEOL JEM-2100F operated at 200 keV.

*2.3. Finite element simulations*

The hydrostatic pressure distribution throughout the α-quartz pillars during compression without e-beam irradiation was investigated using ABAQUS/Standard (Ver. 6.10). We constructed a two-dimensional (2D) axisymmetric model with four-node bilinear axisymmetric quadrilateral elements (CAX4) and adaptive meshing. The deformation behavior of the α-quartz pillar was assumed to be linearly elastic and anisotropic. The six elastic constants $C_{11}$, $C_{33}$, $C_{12}$, $C_{13}$, $C_{14}$, and $C_{44}$ were 86.6, 106.4, 6.74, 12.4, 17.8, and 58.0, respectively [17].

*2.4. Density functional theory simulation*

First-principles molecular dynamics (FPMD) simulations were conducted using the Vienna Ab Initio Simulation Package (VASP, Universität Wien, Wien, Austria) with the PBE functional. We adopted an energy cutoff of 500 eV and used a $2 \times 2 \times 2$ *k*-point mesh. For the simulation structure, we prepared a crystalline quartz cell and generated excess electrons by introducing extra electrons. The same amount of homogeneous background charge of the opposite sign was applied to maintain charge neutrality. The number of excess charges introduced into the model is equal to the number of primary and secondary electrons obtained from the Monte Carlo simulation in which the electron-matter interaction under the irradiation of e-beam with the acceleration voltage of 5 kV and the current density of 21.88 or 87.54 A/m$^2$ was simulated. Considering the timescale accessible with FPMD, we performed the simulation under elevated temperature and pressure: the α-quartz structures were heated to 2500 K in the NVT ensemble and then switched to the NPT ensemble at 2500 K and 35 GPa for 20 ps. Subsequently, the α-quartz structures were quenched at 300 K for 20 ps and then relaxed at 0 K and 0 GPa.

## 3. Results and Discussion

First, we experimentally investigated the deformation of the α-quartz with and without e-beam irradiation. Specifically, we fabricated single-crystalline α-quartz micropillars intentionally along the axis parallel to (0001) and conducted *in situ* uniaxial compression tests. With this axial orientation and loading direction of the pillar, the slip systems of α-quartz, such as basal slip $(0001) < \bar{1}2\bar{1}0 >$ and prismatic slip $(10\bar{1}0) < 0001 >$, cannot be activated because the resolved shear stresses on the slip systems are much lower than the critical resolved shear stress of α-quartz. Instead, we expected that the elastically stored energy from the external loading would be released by the solid-state amorphization, which is accompanied by a volume reduction [14–16]. The pillars were fabricated on a commercial-grade (0001) α-quartz single-crystal substrate through focused-ion-beam milling, and, subsequently, displacement-controlled compression tests were performed on the pillars.

During compression without e-beam irradiation, the α-quartz pillars exhibit a small permanent deformation. Figure 1(a) shows the engineering stress–strain curve of the α-quartz pillar without e-beam irradiation (black curve). As shown, the stress linearly increased with strain, but the curve started to deviate with an incremental decrease at a strain of 0.125. After unloading, the plastic strain of the α-quartz pillar was 0.05. Compared with the uncompressed pillar (Figure 1(b)), the compressed pillar showed a squashed top part (Figure 1(c)), suggesting that a small permanent deformation occurred within it. Note that there is no clear dislocation slip step at the single-crystalline pillar surface. This indicates that plastic deformation did not occur by dislocation movement but by another mechanism; this can be investigated by TEM observation because slip steps are typically generated on the surface along the slip planes as dislocations escape single-crystalline materials [18].

On e-beam irradiation, the α-quartz pillar shows a completely different deformation behavior under compression: large plastic deformation and ductility. In the previous study, we demonstrated that the e-beam-induced deformation behavior of amorphous silica strongly depends on the electron–matter interaction volume [13]. Accordingly, based on the electron–matter interaction model (CASINO™ software (Ver. 3.3)), the acceleration voltage and current density of the e-beam were set to 5 kV and 21.88 A/m$^2$, respectively, ensuring a large electron–matter interaction volume in the α-quartz pillar. The green curve in Figure 1(a) is the engineering stress–strain curve of an α-quartz pillar compressed under e-beam irradiation. As shown, the stress increased linearly up to a strain of 0.07, with a much lower slope compared to that without e-beam irradiation. Subsequently, the slope gradually decreased, and the stress reached its maximum value of 2.94 GPa at a strain of 0.08. Up to this strain, the α-quartz pillar was only compressed along the loading direction, maintaining its straight geometry without localized deformation, as shown in Figure 1(d) (the original shape of the pillar is indicated by a white dashed line). At a strain of 0.08, the stress decreased and plateaued (1.5 GPa). At this stage, the α-quartz pillar deviated from its straight geometry as a result of local transverse deformation (marked with white arrows in Figure 1(d)). After compression, the plastic strain was 0.14, 2.8-times greater than that without e-beam irradiation. The SEM image of the compressed pillar in Figure 1(d) indicates that the α-quartz pillar deformed to a barrel-like shape upon compression under e-beam irradiation. Moreover, because there is no dislocation slip step on the pillar surface, this large plastic deformation is not solely a result of dislocation movement.

Subsequent analysis of the microstructural evolution at different engineering strains enabled us to understand the deformation behavior of the α-quartz pillars under compression during e-beam irradiation. For this analysis, we sampled cross sections of the α-quartz pillars parallel to the direction of the incident e-beam. This revealed that the permanent deformation of pillars with and without e-beam irradiation stems from the solid-state mechanical amorphization. In more detail, Figures 2(a, b) show cross-sectional dark-field TEM images of an α-quartz pillar before and after compression without e-beam irradiation. Dark-field TEM images and selective area diffraction patterns from the $(11\bar{2}0)$ planes clearly show the crystalline (bright regions) and amorphous (dark regions) phases inside the α-quartz pillar. It should be noted that the dark-field TEM image before compression shown in Figure 2(a) shows that the crystalline α-quartz is covered by a 30-nm-thick amorphous layer. This amorphous layer was generated by radiation damage during focused-ion-beam milling to fabricate the micropillars [18]. After compression, the thickness of the amorphous phase in the upper part of the pillar increased to 100 nm (Figure 2(b)). Because there was no additional energy source to increase the temperature, it clearly signifies that the mechanical solid-state amorphization occurred in the top part of the α-quartz pillar during uniaxial compression without e-beam irradiation.

The solid-state mechanical amorphization occurs when α-quartz is subjected to a high hydrostatic pressure of 14 GPa [14–16]. To investigate whether the top part of the pillar experienced sufficient hydrostatic pressure for mechanical amorphization during compression, we constructed a 2D axisymmetric model using ABAQUS/Standard (Ver. 6.10). The calculated engineering stress–strain curve is consistent with that obtained experimentally (gray dashed curve in Figure 1(a)). In particular, the calculated results confirm that the α-quartz pillar deforms elastically to 0.125 strain. Notably, at a strain of 0.125, a hydrostatic pressure of 15 GPa occurred in the α-quartz pillar (Figure 2(c)). Moreover, the region with a hydrostatic pressure above 14 GPa is remarkably similar to the amorphous region of the α-quartz pillar shown in Figure 2(b). Thus, the upper part of the α-quartz pillar was amorphized after elastic deformation owing to the high hydrostatic pressure resulting from compression.

In contrast, the α-quartz pillar under e-beam irradiation was amorphized under a much lower external load and in a much larger region than in the case without e-beam irradiation, and, subsequently, this amorphized region underwent viscoplastic deformation during compression, as observed in our previous study. Next, we captured snapshots of the changes in the deformation behavior and microstructure during e-beam irradiation and compression at engineering strains of 0.04, 0.08, 0.12, and 0.15. Figures 2(d, e) show the corresponding engineering stress–strain curves and dark-field TEM images of the four compressed pillars. The dark-field TEM image in Figure 2(e) shows that the upper part of the α-quartz pillar was amorphized after compression. For the α-quartz pillar compressed to 0.04 engineering strain, the most permanent deformation occurred in the upper part of the pillar. For the α-quartz pillar compressed to an engineering strain of 0.08 in which the engineering stress reached its maximum, the taper angle decreased from 2.4° to 1° after the compression test (Figure S1), suggesting that plastic deformation was no longer restricted to the upper part of the pillar. The corresponding dark-field TEM image in Figure 2(e) shows a significant increase in the amount of amorphous phase in the α-quartz pillar. It should be noted that the amorphized region of the α-quartz pillar compressed to 0.08 strain during e-beam irradiation is already larger than that compressed to 0.15 strain without e-beam irradiation, and almost 40% of the cross-sectional area of the α-quartz pillar was amorphized by compression. Within the pillar, the size of the amorphized region increases nearer to the irradiated surface. Because the temperature increase arising from e-beam irradiation is negligible [8,13], this result suggests that e-beam irradiation facilitates the solid-state mechanical amorphization.

For the α-quartz pillars compressed to 0.12 strain where the engineering stress decreased from its maximum value and reached a plateau, the pillar shape deviated from its original straight geometry (shown by arrows in Figure S1). The TEM image in Figure 2(e) shows that the most permanent deformation occurred in the amorphized region but not in the crystalline region. For the α-quartz pillars compressed to a maximum strain of 0.15 where the stress plateaued, the amorphized area is almost the same as that in the pillar compressed to 0.12 strain. Instead, as shown in the corresponding dark-field TEM image, a large permanent shear deformation occurred in the localized amorphized region. This can be attributed to the viscoplastic deformation of amorphous silica induced by e-beam irradiation [3–8,13]. In our study, the bottom part of the α-quartz pillar that maintains crystallinity acts as a constraint such that the viscoplastic deformation of the amorphized region results in a deformed pillar shape that is significantly different from the original straight shape. The *in situ* experiments and microstructural analysis suggest that, on e-beam irradiation, the top and beam-irradiated surfaces of the α-quartz pillar gradually lose their crystallinity with increasing strain. After the maximum engineering stress, the amorphized region no longer increases, and the viscoplastic deformation of the amorphous phase dominates the entire deformation, leading to the decrease and plateau of the stress.

To understand the mechanism behind the e-beam-enhanced mechanical amorphization of α-quartz theoretically, we performed FPMD simulations using VASP. In our previous work, we found that viscoplastic deformation occurs within the volume where the electron–matter interaction generates excited secondary electrons (or holes) in the material [13]. Specifically, the primary and secondary excited electrons (or holes) can change the nature of the bonds in amorphous silica. Accordingly, we investigated how the primary and secondary electrons affect the amorphization of α-quartz by separately constructing supercell models under hydrostatic pressure with and without these electrons. The irradiation of dielectric materials with a non-penetrating e-beam (such as the e-beam of the SEM in this study) leads to the buildup of primary electrons and secondary excited electrons (or holes) in the material [12,19]. Therefore, we first modeled the excited electron states by introducing additional electrons into the conduction bands in the ground state of the supercell models. The number of excess charges introduced into the model was equal to the number of primary and secondary electrons calculated from Monte Carlo simulation in which a 5 kV e-beam at a current density of 21.88 A/m$^2$ is irradiated. In α-quartz, the basic building block is the [SiO$_4$]$^{2-}$ tetrahedron, as in most silicate materials. Although strong Si–O covalent bonds are formed when the corner oxygen atoms are shared in α-quartz, it is well known that the Si–O bonds also have an ionic nature because of the high electronegativity of the O atom [20]. Consequently, when the amorphization of α-quartz is accompanied by the interpenetration of the [SiO$_4$]$^{2-}$ tetrahedron [16], the Si cations at the center of the tetrahedra result in diverse mid- and long-range configurations in response to the repulsion between adjacent Si cations. Hence, to observe the disorder between the [SiO$_4$]$^{2-}$ tetrahedra, we focused on the Si–Si pair radial distribution function (RDF).

At 2500 K and 35 GPa, the α-quartz exhibited a crystalline-to-amorphous transition. Figure 3 shows how the amorphization behavior changes in the presence of excess electrons. The initial RDF plot of α-quartz at 0 K and 0 GPa (top in Figure 3(a)) shows sharp, intense peaks owing to the periodic atomic arrangement (Figure 3(b)). In contrast, at 2500 K and 35 GPa (middle in Figure 3(a)), the atomic structure of α-quartz without excess electrons (undoped) is distorted, leading to the broadening, splitting, and even disappearance of the original peaks in the RDF plot (black curve). Notably, in α-quartz with excess electrons (doped), at 2500 K and 35 GPa, the broadening and splitting of the first peak are more severe, and the original peak observed at >5 Å disappeared (dark green curve), suggesting that the structure of the doped α-quartz under these conditions is more disordered than that of undoped α-quartz. Moreover, for the undoped α-quartz, after quenching and relaxation at zero external pressure (bottom in Figure 3(a)), the original RDF peaks were observed at the same distances as in the initial state, albeit with decreased intensities, suggesting that the periodicity of the atomic structure was recovered. However, for the doped α-quartz, after quenching and relaxation, no distinct peaks were observed in the RDF plot, except for the first peak, suggesting that the atomic structure remaineds amorphous-like. The atomic configurations shown in Figure 3(b) also show the ordered and disordered structures of undoped and doped α-quartz, respectively, after quenching and relaxation.

The changes in the atomic configuration and the corresponding RDF plot intensified as the number of excess charges increased. The light-green curves in Figure 3(a) indicate the RDF plots of 4×doped α-quartz (corresponding to the 5 kV, 87.54 A/m$^2$ e-beam). At 2500 K and 35 GPa, the RDF plot of the 4×-doped α-quartz showed much broader peaks than that of the undoped quartz, as well as a small shoulder at 2.35 Å. The shoulder position is identical to the Si–Si bond length [21]. In undoped α-quartz, a close Si–Si configuration is not favorable because of the ionic character of the Si–O bond and repulsive force between Si atoms, suggesting that the excess electrons enable severe lattice distortion. After quenching and relaxation, this shoulder remained in the RDF plot, indicating that Si-Si bonds are newly generated owing to the severe lattice distortion in the presence of excess electrons. The corresponding atomic configurations in Figure 3(b) also show the highly disordered structures of 4×-doped α-quartz. The considerable difference in the atomic structures of doped and undoped α-quartz originates from the change in the nature of the interatomic bonding caused by the excess charges. Figure 4(a) shows the density-of-states plots of the doped α-quartz before and at the initial stage of deformation (1 ps). As shown, when excess electrons were introduced into α-quartz, they induced a shift in the Fermi energy level and occupy the conduction band, particularly the O-s, O-p, and Si-s orbitals, which were empty in the undoped α-quartz (top in Figure 4(a)).

The atomic configurations with charge differences shown in Figure 4(b) reveal interesting changes in the charge distribution during deformation. Initially, the excess electrons in the conduction band were evenly distributed around the Si and O atoms at 0 K and 0 GPa (top). However, when the doped α-quartz was subjected to 2500 K and 35 GPa, the density-of-states plot showed defect states in the bandgap generated by the disorder in the atomic structure (1 ps, bottom in Figure 4(a)). The atomic configurations with charge differences at 2500 K and 35 GPa in Figure 4(b) show that excess electrons are transferred to Si atoms having adjacent Si atoms (1 ps, middle) or Si atoms with dangling bonds (3 ps, bottom).

We hypothesize that the excess charges reduce the repulsive force in the distorted atomic structure. Figure 5 illustrates the role of the excess charges in the solid-state mechanical amorphization of α-quartz. Figure 5(a) shows the original tetrahedral $[SiO_4]^{2-}$ configuration along the <0001> direction before deformation. As shown, the $[SiO_4]^{2-}$ tetrahedra form six-membered rings (left-hand image in Figure 5(b)). Under hydrostatic pressure, the Si–O–Si angle changes significantly, leading to the interpenetration of the tetrahedra forming the six-membered ring (right-hand image in Figure 5(b)). Here, interpenetration refers to newly formed close Si–Si or O–O configurations and a loss of periodicity. Owing to the high electronegativity of the O atom, there is a strong repulsive force between the adjacent Si–Si or O–O atoms. The excess charges are expected to be located between these atomic configurations, thus reducing the repulsive force and enabling highly disordered atomic structures, which are unstable within undoped α-quartz (Figure 5(c)). In α-quartz, $[SiO_4]^{2-}$ tetrahedra share an O atom, which may act as a hinge when the atomic structure is distorted. That is, the distortion of the atomic structure that results in the penetration of $[SiO_4]^{2-}$ tetrahedra is more likely to induce the close arrangement of Si atoms rather than O atoms. Therefore, it can be deduced that the positive repulsive force between close Si atoms, which may be a threshold barrier for mechanical amorphization, is alleviated by the excess electrons rather than excess holes. Figure S2 shows that excess holes are located between adjacent O–O atoms at 2500 K and 35 GPa. Notably, the RDF and the corresponding atomic structure indicate that the excess holes induce a disordered atomic structure, although to a lesser extent than the excess electrons. This phenomenon is consistent with the results of Johnson *et al.*, who reported that electron anions act as a type of network modifier that introduces highly mobile weak links into calcium aluminate glass networks, thus complementing the conventional substitutions of anions or cations to lower the glass temperature [22]. Therefore, we suggest that the excess charges introduced by e-beam irradiation behave like localized mobile ions that can lower the high-energy barrier required to achieve a disordered atomic configuration in α-quartz and act as network modifiers, thus facilitating the solid-state amorphization. The same mechanism can be applied to explain the e-beam-induced viscoplastic deformation of amorphous silica observed in previous studies because amorphous silica also consists of a short-range order collective network of $[SiO_4]^{2-}$ tetrahedra.

The interaction volume where excess charges are introduced and generated increases with increase in the current density of the e-beam. Hence, the dependency of the solid-state amorphization of α-quartz on the number of excess charges can be investigated by conducting additional uniaxial *in situ* compression tests under e-beams with different current densities (5 kV and 1.37, 5.47, and 87.54 A/m$^2$). Figure 6(a) shows the engineering stress–strain curves of each compression test to a maximum strain of 0.15 (the curves without e-beam irradiation and with 5 kV, 21.88 A/m$^2$ e-beam irradiation are identical to those in Figure 1). All curves corresponding to samples subjected to e-beam irradiation reached the maximum engineering stress, followed by a decrease and plateau. As the current density increased, the slope of each curve, maximum engineering stress, and corresponding engineering strain all decreased. Furthermore, the plastic strain continued to increase with increase in current density. The SEM images in Figures 6(b–d) show that the compressed shape gradually changed to a barrel-like form as the current density increased. As expected, the corresponding dark-field TEM images (right-hand images in Figure 6(b–d)) indicate an increase in the amorphized area with increase in current density and viscoplastic deformation in it during uniaxial compression as expected.

## 4. Conclusion

In summary, we investigated the deformation behavior of single-crystalline α-quartz micropillars under uniaxial compression during e-beam irradiation. Our *in situ* SEM compression tests revealed that a large permanent deformation occurred in the pillars under e-beam irradiation. Microstructural analysis explained that the deformation originated from the solid-state mechanical amorphization of α-quartz and subsequent viscoplastic deformation of the amorphized region facilitated by e-beam irradiation. In particular, the solid-state amorphization preferentially occurred within the volume where the electron–matter interaction occurred in the α-quartz pillar during compression. Atomic-scale simulations suggested that the e-beam-induced behavior originated from the excess charges that serve as network modifiers, decreasing the repulsive force between Si–Si and O–O atomic configurations and reducing the deformation barrier to the interpenetration of the $[SiO_4]^{2-}$ tetrahedra. We believe that the findings of this study deepen the understanding of the role of excess charges in materials and the electron–matter interaction in brittle oxide materials. Further, they are expected to aid in the development of new methods of glass formation and processing without high temperature and compositional changes.

# Acknowledgements

This study was supported by National Research Foundation of Korea (NRF) grants funded by the Ministry of Science (NRF-2019R1A2C2003430, 2020R1A5A6017701, 2020R1A6A3A03039038, and 2021R1A2C3005096) and by the Creative-Pioneering Researchers Program through Seoul National University. The research facilities at the Institute of Engineering Research at Seoul National University were also utilized for this study.

# Figures

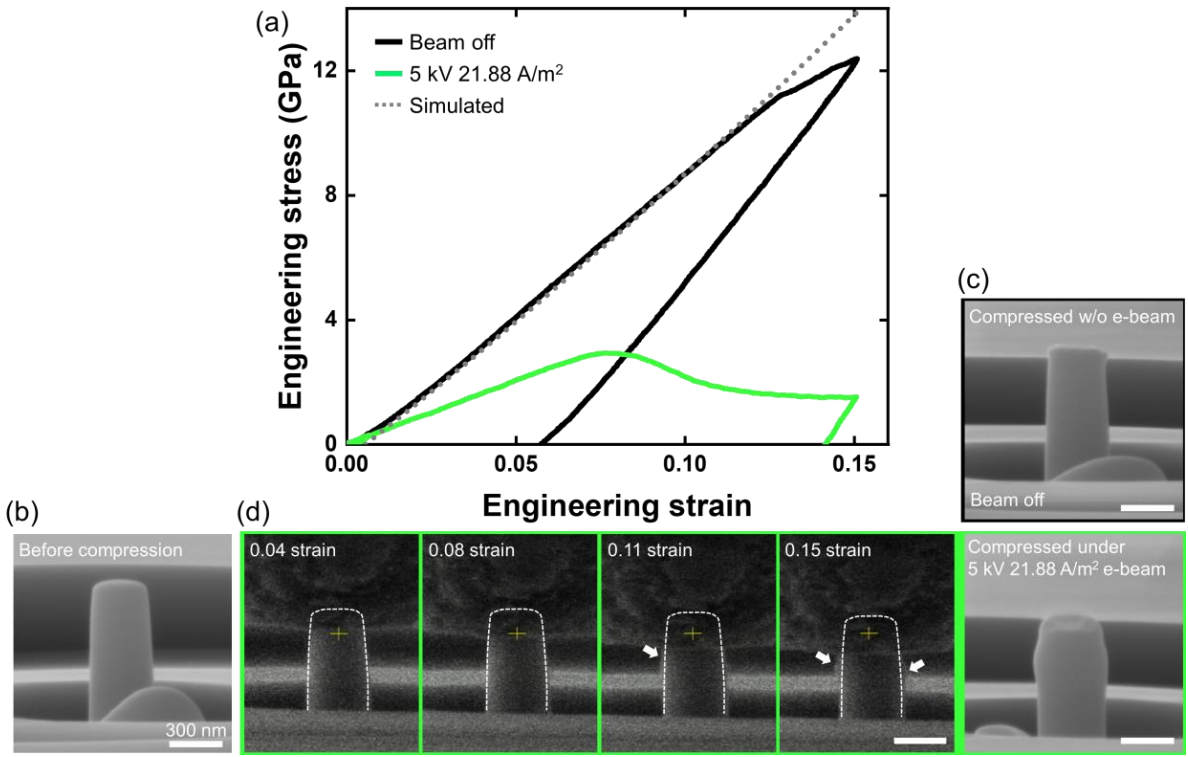

**Figure 1.** *In situ* SEM compression tests on α-quartz micropillar with and without e-beam (5 kV, 21.88 A/m$^2$). (a) Engineering stress–strain curves of micropillars with (green) and without (black) e-beam irradiation. The gray dashed line is a simulated strain–stress curve of a micropillar. (b-d) SEM images of micropillars (b) before compression and (c) after compression without e-beam irradiation, and (d) snapshots of a micropillar compressed under e-beam irradiation. The white dashed line indicates the original shape of the micropillar. White arrows indicate transverse deformation.

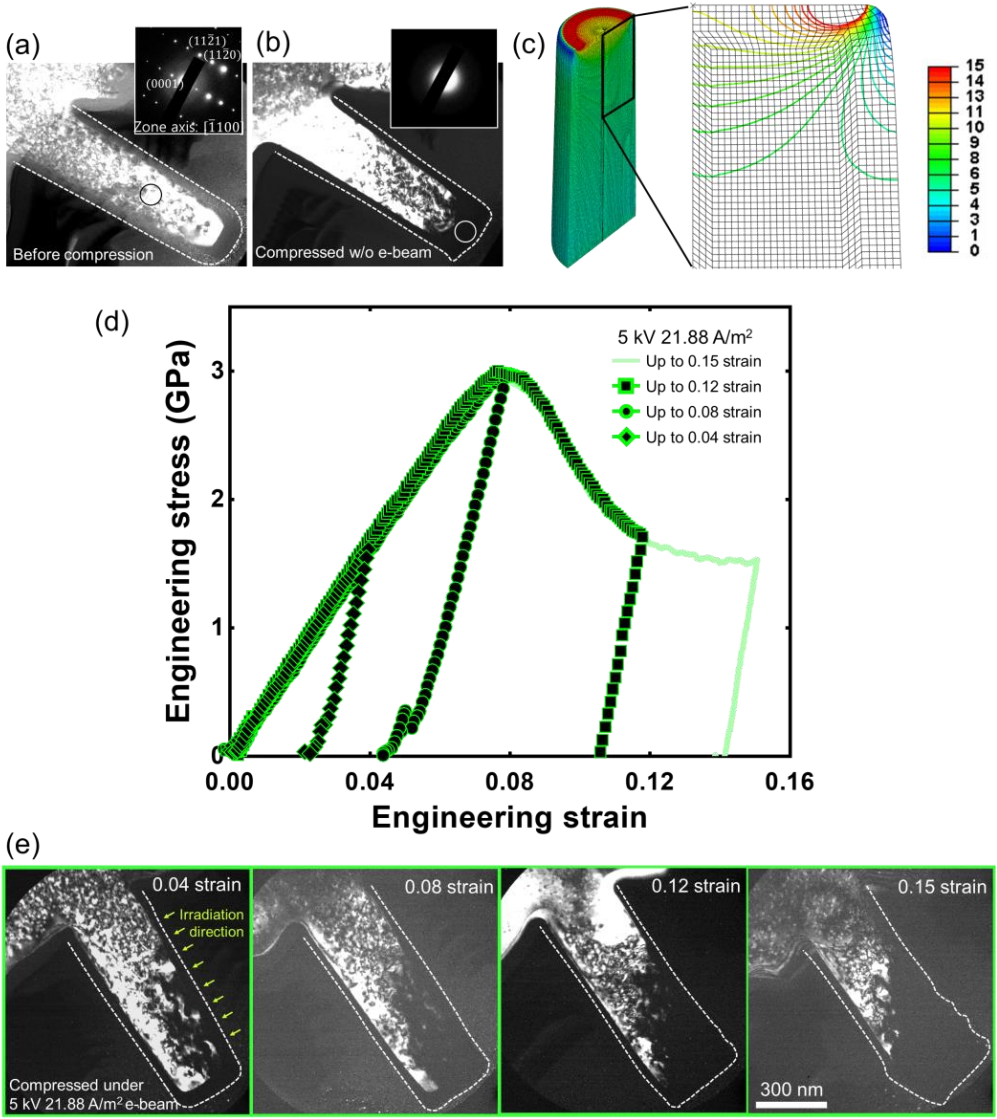

**Figure 2.** Microstructure of α-quartz micropillars. Dark-field TEM images along the $(11\bar{2}0)$ direction of the micropillar (a) before compression and (b) after compression without e-beam irradiation. (c) Hydrostatic pressure distribution in GPa within the pillar at a strain of 0.12. (d) Engineering stress–strain curves of micropillars compressed to strains of 0.04, 0.08, 0.12, and 0.15 under a 5 kV, 21.88 A/m$^2$ electron beam. (e) Dark-field TEM images along the $(11\bar{2}0)$ direction of the micropillars compressed with e-beam irradiation. Green arrows indicate the irradiation direction of the e-beam.

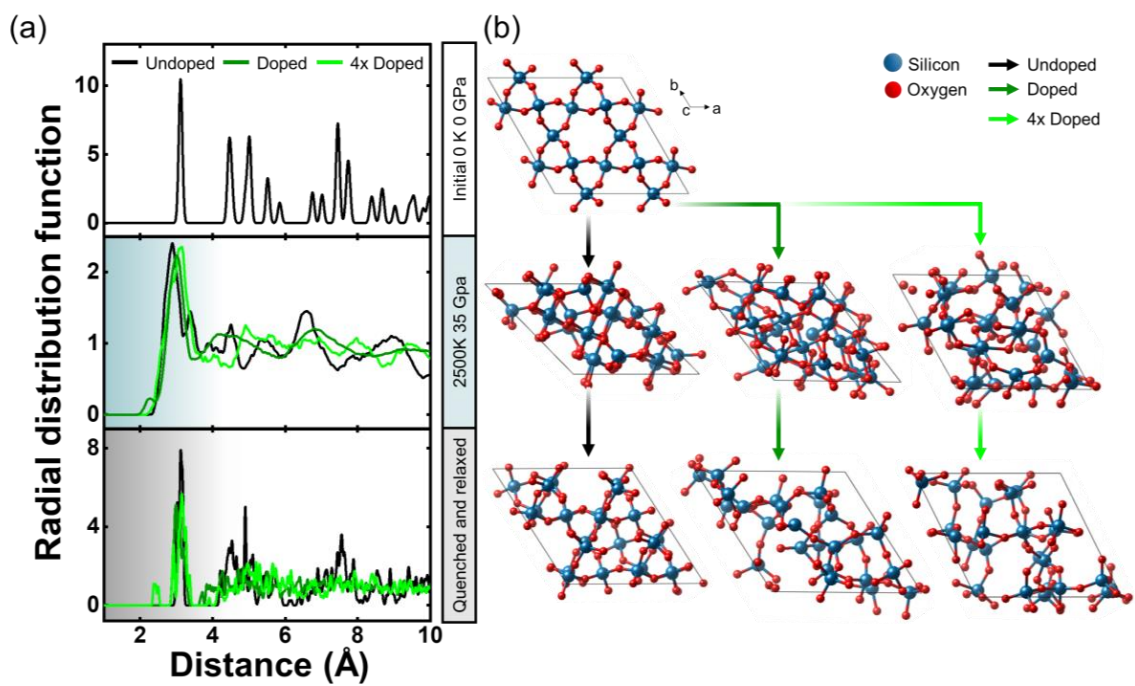

**Figure 3.** Results of density functional theory simulations of α-quartz under hydrostatic pressure with and without excess electrons. (a) RDF plot of a Si–Si pair in α-quartz. (b) Corresponding atomic configurations of α-quartz. Blue and red circles indicate Si and O atoms, respectively.

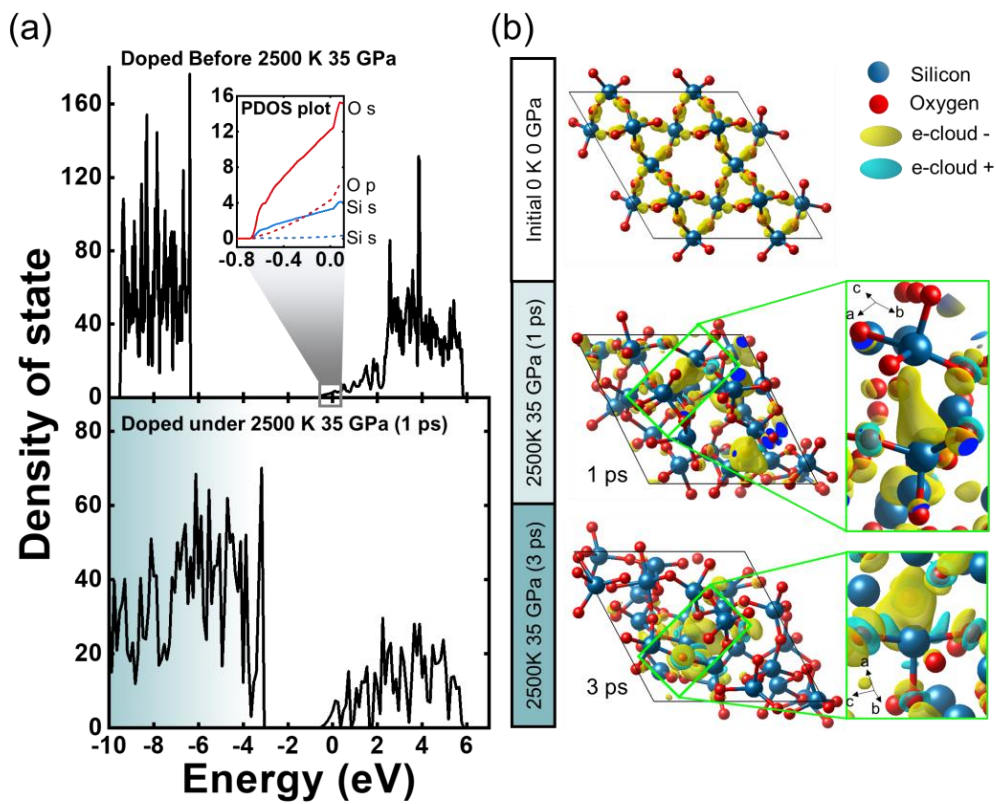

**Figure 4.** Results of density functional theory simulation of α-quartz under hydrostatic pressure with excess electrons. (a) Density-of-states plot of α-quartz with excess electrons. (b) Atomic configurations under hydrostatic pressure showing charge differences. Yellow and blue clouds indicate excess electrons and holes, respectively.

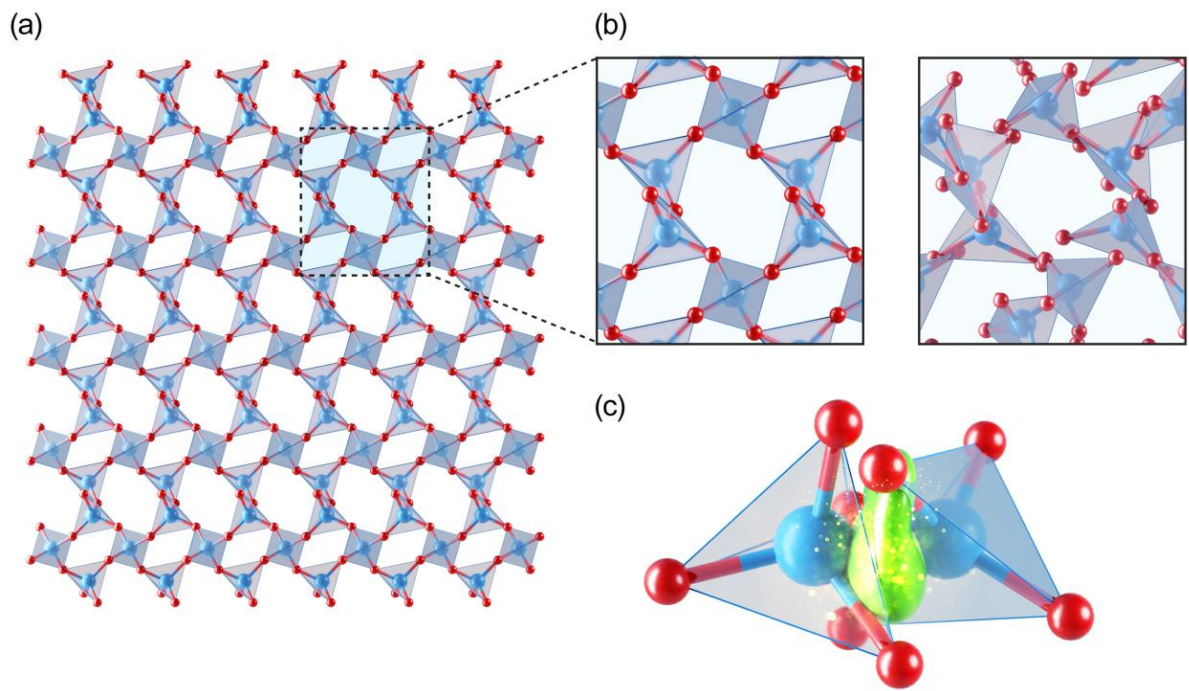

**Figure 5.** Schematic of the solid-state mechanical amorphization of α-quartz and role of excess electrons. (a) Atomic configuration of α-quartz. (b, left) Close-up of $[SiO_4]^{2-}$ tetrahedra forming a six-membered ring. (b, right) Distortion of the six-membered ring and interpenetration of $[SiO_4]^{2-}$ tetrahedra under hydrostatic pressure. (c) Excess electrons located between the Si atoms during interpenetration of $[SiO_4]^{2-}$ tetrahedra.

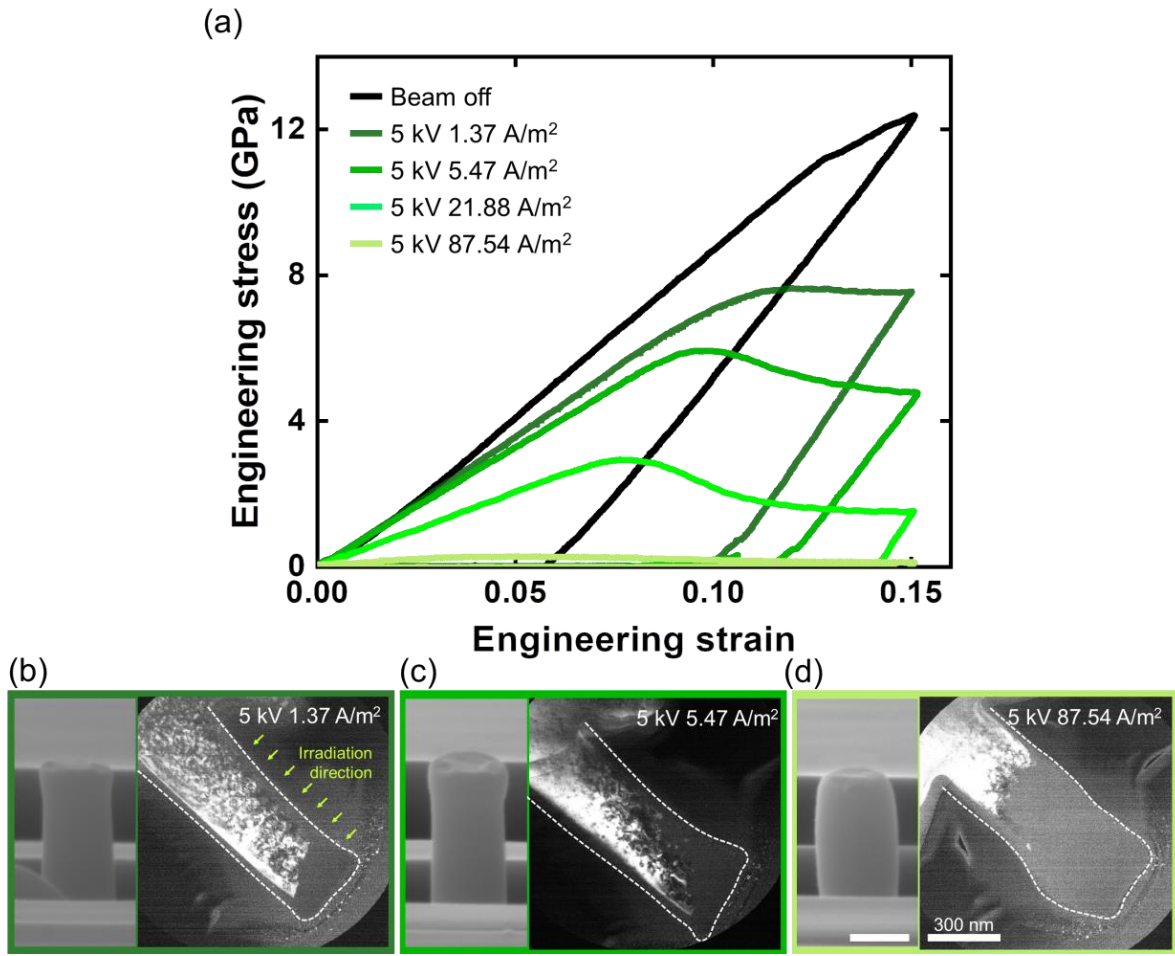

**Figure 6.** *In situ* SEM compression tests on the α-quartz micropillar under various e-beam conditions. (a) Engineering stress–strain curves of micropillars. (b) SEM and $(11\bar{2}0)$ dark-field TEM images of the micropillars.